\documentclass[a4paper, 11pt]{article}

\usepackage[utf8x]{inputenc}
\usepackage{epsfig,amssymb,euscript,xspace, color}
\usepackage{jheppub}
\usepackage[T1]{fontenc}
\usepackage{color}
\usepackage{amssymb}
\usepackage{amsfonts,amsmath,bm}
\usepackage{graphicx}
\usepackage{mathtools}
\usepackage{ragged2e}

\usepackage{dcolumn}
\usepackage{graphicx}
\usepackage{braket}
\usepackage{verbatim}
\usepackage[english]{babel}
\usepackage{hyperref}
\hypersetup{
citecolor=red,
colorlinks=true,
filecolor=red,
linkcolor=blue,
linktocpage=true,
urlcolor=blue
}
\usepackage[titletoc,toc,title]{appendix}
\numberwithin{equation}{section}
\usepackage{cleveref}
\usepackage{epsfig}
\usepackage{float}
\usepackage{amsfonts}
\usepackage{enumitem}
\usepackage[font={footnotesize,it}]{caption}
\usepackage{cases}

\newcommand{\be}{\begin{equation}} 
\newcommand{\ee}{\end{equation}}
\newcommand{\eq}[1]{(\ref{#1})}
\newcommand{\bit}{\begin{itemize}}  \newcommand{\eit}{\end{itemize}}
\newcommand{\ben}{\begin{enumerate}}  \newcommand{\een}{\end{enumerate}}

\newcommand{\rf}[1]{(\ref{#1})}

\def\bd{\begin{document}}
\def\ed{\end{document}}
\def\bea{\begin{eqnarray}}
\def\eea{\end{eqnarray}}
\let\bm=\bibitem

\def\la{\langle}
\def\ra{\rangle}

\def\npb#1#2#3{Nucl. Phys. {\bf{B#1}} #3 (#2)}
\def\plb#1#2#3{Phys. Lett. {\bf{#1B}} #3 (#2)}
\def\prl#1#2#3{Phys. Rev. Lett. {\bf{#1}} #3 (#2)}
\def\prd#1#2#3{Phys. Rev. {D bf{#1}} #3 (#2)}
\def\cmp#1#2#3{Comm. Math. Phys. {\bf{#1}} #3 (#2)}
\def\cqg#1#2#3{Class. Quantum Grav. {\bf{#1}} #3 (#2)}
\def\nppsa#1#2#3{Nucl. Phys. B (Proc. Suppl.) {\bf{#1A}}#3 (#2)}
\def\ap#1#2#3{Ann. of Phys. {\bf{#1}} #3 (#2)}
\def\ijmp#1#2#3{Int. J. Mod. Phys. {\bf{A#1}} #3 (#2)}
\def\rmp#1#2#3{Rev. Mod. Phys. {\bf{#1}} #3 (#2)}
\def\mpla#1#2#3{Mod. Phys. Lett. {\bf A#1} #3 (#2)}
\def\jhep#1#2#3{J. High Energy Phys. {\bf #1} #3 (#2)}
\def\atmp#1#2#3{Adv. Theor. Math. Phys. {\bf #1} #3 (#2)}

\def\sst{\scriptscriptstyle}
\def\thetabar{\bar\theta}
\def\Tr{{\rm Tr}}
\def\one{\mbox{1 \kern-.59em {\rm l}}}

\def\a{\alpha}      \def\da{{\dot\alpha}}  \def\dA{{\dot A}}
\def\b{\beta}       \def\db{{\dot\beta}}
\def\g{\gamma}  \def\G{\Gamma}  \def\dc{{\dot\gamma}}
\def\d{\delta}  \def\D{\Delta}  \def\ddt{\dot\delta}
\def\e{\epsilon}
\def\ve{\varepsilon}
\def\uve{\upvarepsilon}
\def\f{\phi}    \def\F{\Phi}    \def\vvf{\f}
\def\vphi{\varphi}
\def\h{\eta}
\def\k{\kappa}
\def\l{\lambda} \def\L{\Lambda}
\def\m{\mu} \def\n{\nu}
\def\o{\omega}
\def\p{\pi} \def\P{\Pi}
\def\r{\rho}
\def\s{\sigma}  \def\S{\Sigma}
\def\t{\tau}
\def\th{\theta} \def\Th{\Theta} \def\vth{\vartheta}
\def\X{\Xeta}
\def\z{\zeta}

\def\na{\nabla}

\def\cA{{\cal A}} \def\cB{{\cal B}} \def\cC{{\cal C}}
\def\cD{{\cal D}} \def\cE{{\cal E}} \def\cF{{\cal F}}
\def\cG{{\cal G}} \def\cH{{\cal H}} \def\cI{{\cal I}}
\def\cJ{{\mathscr J}} \def\cK{{\cal K}} \def\cL{{\cal L}}
\def\cM{{\cal M}} \def\cN{{\cal N}} \def\cO{{\cal O}}
\def\cP{{\cal P}} \def\cQ{{\cal Q}} \def\cR{{\cal R}}
\def\cS{{\cal S}} \def\cT{{\cal T}} \def\cU{{\cal U}}
\def\cV{{\cal V}} \def\cW{{\cal W}} \def\cX{{\cal X}}
\def\cY{{\cal Y}} \def\cZ{{\cal Z}}
\def\ct{{\cal t}}

\def\ua{\underline{\alpha}}
\def\uc{\underline{\phantom{\alpha}}\!\!\!\gamma}
\def\um{\underline{\mu}}
\def\ud{\underline\delta}
\def\ue{\underline\epsilon}
\def\una{\underline a}\def\unA{\underline A}
\def\unb{\underline b}\def\unB{\underline B}
\def\unc{\underline c}\def\unC{\underline C}
\def\und{\underline d}\def\unD{\underline D}
\def\une{\underline e}\def\unE{\underline E}
\def\unf{\underline{\phantom{e}}\!\!\!\! f}\def\unF{\underline F}
\def\unm{\underline m}\def\unM{{\underline M}}
\def\unn{\underline n}\def\unN{{\underline N}}
\def\unp{\underline{\phantom{a}}\!\!\! p}\def\unP{\underline P}
\def\unq{\underline{\phantom{a}}\!\!\! q}
\def\unQ{\underline{\phantom{A}}\!\!\!\! Q}
\def\unH{\underline{H}}

\def\As {{A \hspace{-6.4pt} \slash}\;}
\def\bs {{b \hspace{-6.4pt} \slash}\;}
\def\Ds {{D \hspace{-6.4pt} \slash}\;}
\def\Gts {{\Gt \hspace{-6.4pt} \slash}\;}
\def\ds {{\del \hspace{-6.4pt} \slash}\;}
\def\ss {{\s \hspace{-6.4pt} \slash}\;}
\def\ks {{ k \hspace{-6.4pt} \slash}\;}
\def\ps {{p \hspace{-6.4pt} \slash}\;}
\def\xs {{x \hspace{-6.4pt} \slash}\;}
\def\pas {{{p_1} \hspace{-6.4pt} \slash}\;}
\def\pbs {{{p_2} \hspace{-6.4pt} \slash}\;}
\def\cFs {{{\cal F} \hspace{-6.4pt} \slash}\;}
\def\Dss {{D \hspace{-7.5pt} \slash}\;}
\def\dss {{\del \hspace{-7.0pt} \slash}\;}

\def\Ah{{\hat{A}}}
\def\Dh{{\hat{D}}}
\def\Gh{{\hat{G}}}
\def\Fh{{\hat{F}}}
\def\Ih{{\hat{I}}}
\def\Jh{{\hat{J}}}
\def\Kh{{\hat{K}}}
\def\Lh{{\hat{L}}}
\def\Ph{{\hat{P}}}
\def\Rh{{\hat{R}}}
\def\Vh{{\hat{V}}}
\def\Xh{{\hat{X}}}

\def\ah{{\hat{\a}}}
\def\bh{{\hat{\b}}}
\def\gh{{\hat{\g}}}
\def\dh{{\hat{\d}}}
\def\rh{{\hat{\r}}}
\def\hh{\hat{h}}
\def\uh{\hat{u}}
\def\xh{\hat{x}}
\def\yh{\hat{y}}
\def\ph{\hat{p}}
\def\xih{\hat{\xi}}
\def\chih{\hat{\chi}}
\def\Psih{\hat{\Psi}}
\def\phih{\hat{\phi}}

\def\psit{\tilde{\psi}}
\def\Psit{\tilde{\Psi}}
\def\Psibt{\tilde{\bar{Psi}}}

\def\lambdat{\tilde {\lambda}}
\def\st{\tilde{\sigma}}

\def\delt{\tilde{\delta}}
\def\Phit{\tilde{\Phi}}
\def\Phitb{\overline{\tilde{Phi}}}
\def\tht{\tilde{\th}}
\def\lt{\tilde{\l}}
\def\chit{\tilde{\chi}}
\def\phit{\tilde{\phi}}

\def\At{\tilde{A}}
\def\Bt{\tilde{B}}
\def\Ct{\tilde{C}}
\def\Dt{\tilde{D}}
\def\Et{\tilde{E}}
\def\Ft{\tilde{F}}
\def\Gt{\tilde{G}}
\def\Ht{\tilde{H}}
\def\It{\tilde{I}}
\def\Jt{\tilde{J}}
\def\Pt{\tilde{P}}
\def\Ot{\tilde{O}}
\def\Mt{\tilde{M }}
\def\Nt{\tilde{N}}
\def\St{\tilde{S}}
\def\Vt{\tilde{V}}
\def\Xt{\tilde{X}}
\def\at{\tilde{a}}
\def\dt{\tilde{d}}
\def\htt{\tilde{h}}
\def\ft{\tilde{f}}
\def\gt{\tilde{g}}
\def\pt{\tilde{p}}
\def\qt{\tilde{q}}
\def\rt{\tilde{r}}
\def\nt{\tilde{n}}
\def\ut{\tilde{u}}
\def\wt{\tilde{w}}
\def\zt{\tilde{z}}
\def\xt{\tilde{x}}
\def\yt{\tilde{y}}
\def\Psit{\tilde{\Psi}}
\def\phit{\tilde{\phi}}
\def\tD{\tilde{\D}}

\def\eb{\bar{\epsilon}}
\def\delb{\bar{\partial}}
\def\thb{\bar{\theta}}
\def\mub{\bar{\mu}}
\def\lamb{\bar{\l}}
\def\psib{\bar{\psi}}
\def\sb{\bar{\sigma}}
\def\xib{\bar{\xi}}
\def\chib{\bar{\chi}}

\def\Psib{\bar{\Psi}}
\def\Phib{\bar{\Phi}}
\def\Lamb{\bar{\Lambda}}
\def\Sb{{\overline \Sigma}}
\def\cb{\bar{c}}
\def\hb{\bar{h}}
\def\qb{\bar{q}}
\def\wb{\bar{w}}
\def\ub{\bar{u}}
\def\zb{{\bar{z}}}
\def\Hb{\bar{H}}
\def\Qb{{\bar Q}}
\def\Omegab{\overline{\Omega}}
\def\ob{\overline{\omega}}

\def\Ab{{\overline A}} \def\Bb{{\overline B}} \def\Cb{{\overline C}}
\def\Db{{\overline D}} \def\Eb{{\overline E}} \def\Fb{{\overline F}}
\def\Gb{{\overline G}}
\def\Ib{{\overline I}}
\def\Jb{{\overline J}} \def\Kb{{\overline K}} \def\Lb{{\overline L}}
\def\Mb{{\overline M}} \def\Nb{{\overline N}} \def\Ob{{\overline O}}
\def\Pb{{\overline P}}  \def\Rb{{\overline R}}
 \def\Tb{{\overline T}} \def\Ub{{\overline U}}
\def\Vb{{\overline V}} \def\Wb{{\overline W}} \def\Xb{{\overline X}}
\def\Yb{{\overline Y}} \def\Zb{{\overline Z}}

\def\fb{{\overline f}}
\def\gb{{\overline g}}
\def\nb{{\overline n}}
\def\mb{{\overline m}}
\def\lb{{\overline l}}
\def\yb{{\overline y}}

\def\ldel{{\overleftarrow{\del}}}
\def\rdel{{\overrightarrow{\del}}}
\def\ldeldel{{\overleftarrow{\del^2}}}
\def\rdeldel{{\overrightarrow{\del^2}}}
\def\ldelb{{\overleftarrow{\bar{\del}}}}
\def\rdelb{{\overrightarrow{\bar{\del}}}}

\def\ba{{\bf a}}
\def\bk{{\bf k}}
\def\bl{{\bf l}}
\def\bp{{\bf p}}
\def\bq{{\bf q}}
\def\br{{\bf r}}
\def\bt{{\bf t}}
\def\bu{{\bf u}}
\def\bv{{\bf v}}
\def\bx{{\bf x}}
\def\by{{\bf y}}
\def\bA{{\bf A}}
\def\bR{{\bf R}}
\def\bV{{\bf V}}

\def\bz{{\boldsymbol{\zeta}}}

\def\bone{{\bf 1}}

\def\va{{\vec a}}
\def\vk{{\vec k}}
\def\vp{{\vec p}}
\def\vq{{\vec q}}
\def\vx{{\vec x}}
\def\vy{{\vec y}}
\def\vu{{\vec u}}
\def\vv{{\vec v}}
\def \vH{{\vec H}}
\def \vg{{\vec g}}

\def\vs{{\vec \sigma}}
\def\vtau{{\vec \tau}}

\newcommand{\ov}[1]{\overrightarrow{#1}}

\def\frA{\mathfrak{A}}
\def\frB{\mathfrak{B}}
\def\frC{\mathfrak{C}}
\def\frD{\mathfrak{D}}
\def\frE{\mathfrak{E}}
\def\frF{\mathfrak{F}}
\def\frG{\mathfrak{G}}
\def\frH{\mathfrak{H}}
\def\frM{\mathfrak{M}}
\def\frN{\mathfrak{N}}
\def\frR{\mathfrak{R}}
\def\frW{\mathfrak{W}}

\def\fra{\mathfrak{a}}
\def\frb{\mathfrak{b}}
\def\frf{\mathfrak{f}}
\def\frg{\mathfrak{g}}
\def\frh{\mathfrak{h}}
\def\frl{\mathfrak{l}}
\def\frs{\mathfrak{s}}
\def\fri{\mathfrak{i}}
\def\frj{\mathfrak{j}}

\def\ma{\mathfrak{a}}
\def\mg{\mathfrak{g}}
\def\mh{\mathfrak{h}}
\def\mR{\mathfrak{R}}
\def\mN{\mathfrak{N}}

\newcommand{\nn}{{\nonumber}}

\def\d{\delta}\def\D{\Delta}\def\ddt{\dot\delta}

\def\pa{\partial} \def\del{\partial}
\def\xx{\times}
\def\uno{\mbox{1 \kern-.59em {\rm l}}}

\def\trp{^{\top}}
\def\inv{^{-1}}
\def\dag{\dagger}
\def\pr{^{\prime}}

\def\rar{\rightarrow}
\def\lar{\leftarrow}
\def\lrar{\leftrightarrow}

\newcommand{\0}{\,\!}      
\def\one{1\!\!1\,\,}
\def\im{\imath}
\def\jm{\jmath}

\newcommand{\tr}{\mbox{tr}}
\newcommand{\slsh}[1]{/ \!\!\!\! #1}

\newcommand{\1}{\mbox{1}\hspace{-0.25em}\mbox{l}}

\def\vac{|0\rangle}
\def\lvac{\langle 0|}

\def\hlf{\frac{1}{2}}
\def\ove#1{\frac{1}{#1}}
\newcommand{\hot}[1]{\frac{#1}{2}}

\def\Box{\square}
\def\CC {\mathbb{C}}
\def\FF {\mathbb{F}}
\def\RR{\mathbb{R}}
\def\NN{\mathbb{N}}
\def\ZZ{\mathbb{Z}}
\def\bb#1{{\bf #1}}
\def\bcomment#1{}
\def\bfhat#1{{\bf \hat{#1}}}
\def\VEV#1{\left\langle #1\right\rangle}

\newcommand{\ex}[1]{{\rm e}^{#1}} \def\ii{{\rm i}}

\newcommand{\lrbrk}[1]{\left(#1\right)}
\newcommand{\lrsbrk}[1]{\left[#1\right]}
\newcommand{\sfrac}[2]{{\textstyle\frac{#1}{#2}}}

\def\stw{{\sqrt{2}}}

\def\rf {{\rm f}}
\def\ri {{\rm i}}
\def\rj {{\rm j}}
\def\rn {{\rm n}}
\def\rk {{\rm k}}
\def\rl {{\rm l}}
\def\rr {{\rm r}}

\def\rQ {{\scriptscriptstyle \rm \cQ}}
\def\rR {{\scriptscriptstyle \rm \cR}}

\def\cQb{{\cal \Qb}}
\def\cRb{{\cal \Rb}}
\def\cWb{{\cal \Wb}}

\def\fd {{\rm N}}
\def\afd {{\overline{\rm N}}}

\def \II {I\hspace{-.1em}I\hspace{.1em}}
\def \IIA {\mbox{\II A\hspace{.2em}}}
\def \IIB {\mbox{\II B\hspace{.2em}}}
\def \gs {g^s}
\def \ls {\lambda^s}

\def \I {{\cal I}}
\def \qs {q\hspace{-.53em}/\hspace{.15em}}
\def \ks {k\hspace{-.53em}/\hspace{.15em}}
\def \YM {{\mbox{\tiny YM}}}
\def \gym {g_{\YM}}

\def \Lc {\L_c}
\def\IR{\relax{\rm I\kern-.18em R}}
\def \id {{\bf 1}}

\def\cci{\ell}
\def\ccj{\ell'}

\def\bbq{\pmb{q}}
\def\bom{\pmb{\o}}
\def\bJ{\pmb{J}}
\def\bM{\pmb{M}}
\def\bB{\pmb{B}}
\def\bn{\pmb{n}}
\def\bE{\pmb{E}}

\title{\Large \bf\boldmath Anomaly Induced Current in Boundary Lifshitz Field Theory}

\author[a,b,c]{Chong-Sun Chu}
\author[b,c,d]{Himanshu Parihar}

 \affiliation[a]{
	Department of Physics, National Tsing-Hua University,
 Hsinchu 30013, Taiwan}
\affiliation[b]{Center for Theory-Computation-Data Science Research,
National Tsing-Hua University,
 Hsinchu 30013, Taiwan}
\affiliation[c]{Physics Division,
    National Center for Theoretical Sciences,
     Taipei 10617, Taiwan}
\affiliation[d]{Department of Physics, College of Science, Kyung Hee University,
Seoul 02447, Republic of Korea}

\emailAdd{cschu@phys.nthu.edu.tw}
\emailAdd{himansp@khu.ac.kr}

\abstract{\noindent We study quantum transport phenomena induced by anisotropic Lifshitz
scale anomaly in a boundary Lifshitz field theory (BLFT) coupled to an
external electromagnetic background. In this context, we obtain the
anisotropic scale anomaly in Lifshitz field theories coupled to a
background $U(1)$ gauge field and subsequently compute the anomaly
induced near boundary current in a BLFT. Focusing on 5D BLFTs, we find
that the temporal and spatial components of the induced current
exhibit distinct power law dependencies on the distance from the
boundary, reflecting the intrinsic time–space anisotropy of the
theory. We further derive this anomalous current holographically from
the bulk dual of BLFT and find that the temporal component is
independent of the boundary conditions while the spatial component
depends explicitly on them.  The distance dependence is in
exact agreement with the dual field theory result. }

\begin{document}
	
	\maketitle
	\flushbottom
	\pagebreak

	\definecolor{orange}{rgb}{1.0, 0.49, 0.0}

\section{Introduction}

The investigation of anomaly induced currents has recently emerged as
a central topic in theoretical physics, offering profound insights
into a wide range of physical phenomena spanning vastly different
energy scales. A prominent example is the chiral anomaly in
non-Abelian gauge theories, which imposes nontrivial constraints on
the fundamental interactions of chiral fermions within the Standard
Model \cite{Peskin:1995ev} (see also
\cite{Kharzeev:2013ffa,Landsteiner:2016led}). Closely related
phenomena include the chiral magnetic effect (CME)
\cite{Vilenkin:1995um,Vilenkin:1980fu, Giovannini:1997eg,
  PhysRevLett.81.3503, Fukushima:2012vr} characterized by the
generation of current which align with the external magnetic field, and the chiral vortical effect (CVE)
\cite{Kharzeev:2007tn,Erdmenger:2008rm,Banerjee:2008th,Son:2009tf,
  Landsteiner:2011cp,Golkar:2012kb,Jensen:2012kj}
where a current is induced by the rotational motion of a charged
fluid.  It has also been shown that anomalous currents can arise in
conformally flat gravitational backgrounds as a consequence of the
Weyl anomaly \cite{Chernodub:2016lbo,
  Chernodub:2017jcp,Chernodub:2018ihb,Chernodub:2019blw}. Since
anomalies are intrinsic properties of the quantum vacuum, an
intriguing question is whether anomalous transport phenomena can occur
in vacuum, even in the absence of any material medium. A primary
example of such vacuum driven physics is the Casimir effect
\cite{Casimir:1948dh,Plunien:1986ca,Bordag:2001qi}, which arises from
the vacuum energy's sensitivity to boundary conditions. The Casimir
effect has been studied in a highly general framework, allowing for
arbitrary boundary geometries and curved spacetime backgrounds. In
this context, universal relations between Casimir coefficients and
boundary central charges in boundary conformal field theories (BCFTs)
have been studied \cite{Miao:2017aba}. This analysis was carried out
using field theoretic methods, focusing on the behavior of the
energy–momentum tensor in the vicinity of boundaries.

The analysis was further extended to
boundary systems possessing a global $U(1)$ symmetry in \cite{Chu:2018ksb},
leading to the
discovery of a novel class of anomalous current localized near the
boundary and generated by the Weyl anomaly. In four dimensions, this
current is induced near the boundary by a background field strength
and is determined universally by the central charge, which is in turn
is related to its beta function. Unlike previously studied anomalous
transport phenomena, this effect arises in the zero-temperature vacuum
of flat spacetime and does not rely on any material medium. Instead,
it represents an intrinsic feature of the quantum vacuum’s sensitivity
to boundaries analogous to the Casimir effect. A key aspect of these
findings is that the anomalous currents originate directly from the
Weyl anomaly associated with background fields. The conformal field
theories defined on manifolds with boundaries
\cite{Cardy:1984bb,McAvity:1993ue,Cardy:2004hm} have long been an
active area of research. A non-perturbative holographic dual
description of boundary conformal field theories (BCFTs) was proposed
in \cite{Takayanagi:2011zk,Fujita:2011fp}, in which the gravitational
dual consists of a portion of AdS spacetime truncated by an end of the
world (EOW) brane. The location of the EOW brane is fixed by boundary
conditions imposed on the bulk gravitational fields. These boundary
conditions may take the form of Neumann boundary conditions, as
originally introduced in \cite{Takayanagi:2011zk,Fujita:2011fp}, or
alternatively conformal boundary conditions
\cite{Miao:2017gyt,Chu:2017aab,Chu:2021mvq} or Dirichlet boundary
conditions \cite{Miao:2018qkc}, as investigated in subsequent
studies. All of these choices of boundary conditions lead to a
consistent realizations of the AdS/BCFT. Building on this framework,
anomalous current transport in holographic BCFTs was investigated in
\cite{Chu:2018ntx}. Interestingly, this current was shown to be
independent of specific boundary conditions in four dimensions, though
it becomes dependent on boundary conditions in higher dimensions. This
holographic approach has since been extended to six-dimensional BCFTs
\cite{Chu:2018fpx,Chu:2019rod}, leading to the prediction of an $N^3$
scaling of degrees of freedom in non-Abelian 2-form gauge theories of
$N$ M5-branes in the large-$N$ limit. Additional consequences of the
Weyl anomaly in spacetimes with boundaries have also been
uncovered. These include the emergence of Fermi Condensation
\cite{Chu:2020mwx,Chu:2020gwq}, anomalous current for conformally flat
spaces with arbitrary scale factors \cite{Zheng:2019xeu}, chiral
currents \cite{Chu:2022bhj}, and a novel vacuum spin transport effect
induced by electromagnetic fields \cite{Chu:2021eae} (see also
\cite{Miao:2018dvm,Hu:2020puq,Liu:2021lbh,Guo:2021dzz,Miao:2022oas}).

In relativistic quantum field theory, scale invariance is known to be
generically enhanced to full conformal invariance. This enhancement,
however does not occur in non-relativistic field theories which has
less constrained symmetry structure. One important example is Lifshitz
scaling symmetry defined by
\begin{equation}
t \to \l^z t, \quad x^i \to \l x^i, \quad \l >1,
\end{equation}
where $z$ denotes the dynamical critical exponent. This symmetry has
proven essential in describing various physical systems, most notably
the Quantum Lifshitz Model (QLM) at $z=2$ in $2+1$ dimensions
\cite{Ardonne_2004}. The QLM serves as a bridge to 2D Conformal Field
Theories (CFTs) \cite{PhysRevB.23.4615} and provides an effective
framework for the Rokhsar-Kivelson dimer model
\cite{PhysRevLett.61.2376}.
From the holographic perspective, Lifshitz
field theories (LFTs) are typically dual to bulk theories of the
Einstein–Proca type, which include a massive vector field
\cite{Kachru:2008yh,Taylor:2008tg,Taylor:2015glc}. This feature
distinguishes them from the asymptotically AdS spacetimes of the
AdS/CFT correspondence, as the Lifshitz bulk geometry also involves an
additional massive gauge field. Such holographic constructions are
particularly valuable because they offer a framework for studying
quantum critical points and renormalization group (RG) flows in
non-relativistic systems. Analogous to
the trace anomalies in relativistic
theories, non-relativistic quantum field theories can exhibit Lifshitz
scale anomalies where classical Lifshitz symmetry is broken by quantum
effects. The Lifshitz scale anomalies in QLM have been analyzed in
\cite{Griffin:2011xs,Baggio:2011ha} where it was shown that the
general form of the anomaly is fixed up to two independent central
charges. These central charges were computed using heat-kernel
techniques in a free Lifshitz scalar theory \cite{Baggio:2011ha} and
independently via holographic renormalization in a dual gravitational
model \cite{Baggio:2011cp}. In addition, cohomological analyses have
been carried out to classify the possible structures of Lifshitz scale
anomalies \cite{Arav:2014goa,Arav:2016xjc,Pal:2016rpz}. Further
developments in this direction can be found in
\cite{Horava:2008ih,Horava:2009uw,Horava:2008jf,Adam:2009gq,Ross:2011gu,Mann:2011hg,
  Gomes:2011di,Griffin:2012qx,Holsheimer:2013ula,Jensen:2014hqa,
  Auzzi:2015fgg,Arav:2016akx,Perez-Nadal:2016tzr,Barvinsky:2017mal,
  Ahmadain:2019vri}.

Recently, the quantum effects of Lifshitz anisotropy on the phenomena of
vacuum tunneling  were studied and applied to the discussion
of Josephson junction\cite{Chu:2025cqq}.
Furthermore, 
a Lifshitz scalar field theory with an arbitrary dynamical exponent
$z$ has been formulated in \cite{Basak:2023otu} (see also
\cite{Benedetti:2023pbt,Benedetti:2024oif} for related work). It was
shown that the Lifshitz-invariant ground state of the theory takes a
special Rokhsar–Kivelson (RK) form
\cite{PhysRevLett.61.2376,Henley_2004}, and its entanglement
properties were analyzed.  Building on this, \cite{Chu:2024nwf}
proposed a holographic dual description of Lifshitz field theories in
the presence of a boundary. In this construction, the bulk geometry is
given by a portion of Lifshitz spacetime truncated by an end of the
world (EOW) brane. It was shown quite surprisingly that, for boundary
Lifshitz field theories (BLFT) defined on a half-space, Neumann and
conformal boundary conditions lead to the same brane profile despite
the anisotropic nature of the bulk geometry. Moreover, two distinct
holographic $g$-functions were introduced which were based on the null
energy condition and the dominant energy condition, imposed on the EOW
brane and these $g$-functions were shown to decrease monotonically
along the boundary RG flow.

The developments discussed above naturally raise the question of
whether similar induced phenomena arise when Lifshitz field theories
are coupled to additional external backgrounds. A particularly
interesting case is that of an electromagnetic (EM) background. In
this work, we investigate quantum transport phenomena induced by
anisotropic Lifshitz scale anomaly in a BLFT coupled to an external EM
field. We begin by
classifying the possible
anisotropic Lifshitz anomaly terms in a
Lifshitz field theory due to the presence of a background $U(1)$ gauge
field. Using this result, we then compute the anomaly induced
current in a boundary Lifshitz field theory. For the specific case of
five dimensional BLFTs, we find that the induced current exhibits
distinct dependence on the distance from the boundary in the temporal
and spatial components, reflecting the intrinsic anisotropy of the
theory. Finally, we provide a holographic derivation of this anomalous
current from the bulk dual description of a BLFT. The holographic
analysis reveals that in six dimensional bulk dual to the five
dimensional LFTs, the temporal component of the current remains
universal and independent of boundary conditions, whereas the spatial
component depends explicitly on them. Furthermore, we demonstrate that
the distance dependence of these holographic currents is in agreement
with the results obtained from the field theory.

The rest of the paper is organized as follows. In \cref{sec-An-Weyl}
we
classify
the anisotropic Lifshitz scale anomaly for Lifshitz
field theory coupled to an external gauge field. In
\cref{sec-current-BLFT} we show that the Lifshitz anomaly leads to an
anomalous current and obtain the anomalous current in 5D BLFTs. Then
in \cref{sec-Hol-currrent}, we first review the construction of the
bulk dual of a BLFT and subsequently derive the holographic current in
dual BLFTs. Finally we conclude with a summary and discussion in
\cref{sec-summary}.

\section{Lifshitz scale anomaly}\label{sec-An-Weyl}

Lifshitz field theories are a class of anisotropic field theories that
preserve time and space separately (foliation-preserving
diffeomorphisms $t\to \tilde{t}(t), x^i\to \tilde{x}^i(x^i,t)$). To
couple such a theory to background fields one uses an ADM/foliation
decomposition of the background geometry with fields ($N(t,x),
N_i(t,x),g_{ij }(t,x)$) in $(d+1)$-dimensions as
\begin{equation}\label{ADM-form}
ds^2 = N^2 dt^2 + g_{ij} dx^i dx^j,
\end{equation}
so that the spacetime volume element is $d^{d+1}x\, N\sqrt{g}$. Note
that the shift function $N^i$ is not included here as it can be
removed locally by a foliation preserving diffeomorphism.  Similar to
the trace anomalies in relativistic field theories, quantum
corrections in non-relativistic field theories can break classical
Lifshitz symmetry leading to the anisotropic Lifshitz scale anomalies.
Under the local anisotropic scale transformation with parameter
$\sigma(t,x)$ and dynamical exponent $z$,
the background fields transform infinitesimally ($\delta\sigma\ll 1$) as
\begin{equation}
\delta_\sigma N = z\,N \,\delta \sigma+ \cdots, \qquad
\delta_\sigma g_{ij} = 2 g_{ij}\,\delta\sigma + \cdots,
\end{equation}
so that $N\sqrt{g}$ transforms with weight $z + d$. Let
$W[N,g_{ij};\{g_a\} ]$ be the renormalized effective action of the
Lifshitz theory coupled to the background fields and couplings $g_a$.
In the presence of an anomaly, the Lifshitz scale anomaly is captured
by the variation of the effective action $W$ and is given by
\begin{equation}\label{Variation-W}
  \delta W = \int d^{d+1}x \, N\sqrt{g}\, \mathcal{I}(t,x)\,\delta\sigma,
\end{equation}
where $\mathcal{I}(t,x)$ is the local Lifshitz scale anomaly density. For a constant (global) parameter $\sigma$, the integrated Lifshitz anomaly $\mathcal{A}$ is given as
\begin{equation}\label{Int-anomaly}
\mathcal{A}= \int d^{d+1}x\, N\sqrt{g}\, \mathcal{I}(t,x).
\end{equation}
Our goal is to identify the possible terms that may appear in the
anomaly density $\mathcal{I}(t,x)$ when the theory is coupled to a
background $U(1)$ gauge field. A systematic cohomological analysis of
Lifshitz scale anomalies in the presence of metric or scalar field
etc., has been carried out in \cite{Arav:2014goa,Arav:2016xjc}
providing a general framework for classifying the allowed anomaly
structures. Since the anomaly density $\mathcal{I}(t,x)$ must be a
local scalar constructed from the background fields e.g. $(N, g_{ij})$,
the admissible terms are subject to strong
constraints. First, dimensional analysis and scaling weight
(canonical mass dimension)
considerations require that $\mathcal{I}$ carry total anisotropic
weight $(z+d)$, ensuring that $N\sqrt{g} \,\mathcal{I}$ is anisotropic
scale invariant up to the overall factor appearing in the integrated
anomaly. Second, $\mathcal{I}$ must be invariant under
foliation-preserving diffeomorphisms including time reparametrizations
and spatial diffeomorphisms. Finally, only cohomologically nontrivial
densities (relative cohomology of the Lifshitz scaling operator with
respect to foliation preserving diffeomorphisms) give rise to genuine
anomalies \cite{Arav:2014goa}. For example, in $(2+1)$-dimensional
theories with $z=2$, the allowed anomaly densities are
spatial-curvature and extrinsic-curvature type scalars built from
$g_{ij},N$ and their spatial derivatives/time derivatives arranged to
have the right weight \cite{Griffin:2011xs,Baggio:2011ha}.

In this paper we are interested in the additional anomalous effect due
to a background $U(1)$ gauge field. 
Since we are considering flat background metric and pure gauge sector: the relevant symmetries are only the global Lifshitz scaling and the $U(1)$ gauge transformations, and these two operations commute i.e $[\delta_\sigma, \delta_\lambda^{\mathrm{gauge}}]=0$. Hence, the Wess–Zumino consistency conditions are automatically satisfied as long
as the Lifshitz anomaly ${\cal I} $ is Lifshitz scale invariant and gauge invariant.
Due to the space time anisotropy, it is necessary to
decompose the background 
gauge field $A_\mu=(A_t,A_i)$ into components that are adapted to the
foliation \eq{ADM-form}. The desired invariants
can then be built from the gauge
field strength components i.e $E_i, F_{ij}$ and their (spatial)
derivatives, chosen to have the correct anisotropic weight so that it
is marginal under the Lifshitz scaling. Unlike relativistic theories
where quadratic expression involving $F_{\mu\nu}$
must be $F_{\mu\nu}F^{\mu\nu}$ 
due to Lorentz invariance, there are more Lifshitz
invariants that can be form from 
the electric field, the magnetic field and their spatial derivatives,
after taking into account of constraints required by
the anisotropic scale invariance.
The analysis of scaling weights are presented in the
appendix \ref{dim}. For example, with $z=2$ in $(4+1)$-dimensions, the
theory is classically scale invariant corresponding to a dimensionless
gauge coupling $([e]=0)$. At the quantum level, however this scale
invariance is generically broken leading to a Lifshitz scale
anomaly. On dimensional grounds, the Lifshitz anomaly density for this
case can therefore be written as
\begin{equation}\label{A-5D}
    \mathcal{I}=c_T\,F_{ti}F^{ti} + c_S\, F_{ij}\,\nabla^2 F^{ij},
\end{equation}
where
the coefficients $c_T$ and $c_S$ are the central charges of the
Lifshitz scale anomaly. The cohomological analysis for \eq{A-5D} has been carried out in appendix \ref{Coh-analysis}.
One may construct mixed curvature gauge invariants such as
$RF_{ij}F^{ij}, R_{ij}F_{k}^iF^{jk}$ or $K F_{ij}F^{ij}$.  Since our
focus is on the pure gauge sector, we do not consider these
mixed anomaly terms
here.  In the following section, we consider the specific case
\eq{A-5D} for $(4+1)$-dimensions with $z=2$.

\section{Anomalous current in boundary Lifshitz field theory}
\label{sec-current-BLFT}

In this section, we demonstrate that for a broad class of boundary
Lifshitz field theories (BLFTs) with $U(1)$ gauge symmetry, the
anisotropic Lifshitz scale anomaly gives rise to a induced current
near the boundary. To derive this anomalous current, we employ the
methods based on the transformation properties of the current under
Weyl rescaling following the approach of \cite{Zheng:2019xeu}. We
begin by deriving the transformation law of the current under
anisotropic Weyl (scale) transformations \eq{anisotropi-Wey-transf} in
a Lifshitz field theory. Using this result, we then obtain the
anomalous current for a BLFT defined on flat space with a planar
boundary by exploiting the relation between Lifshitz spacetime and
flat space through an anisotropic local transformation. This procedure
makes explicit how the breaking of classical Lifshitz scaling at the
quantum level manifests as an induced current near the boundary.
Consider a finite local anisotropic scale transformation
\begin{equation}\label{anisotropi-Wey-transf}
N \mapsto N' = e^{z\sigma} N, 
\qquad
g_{ij} \mapsto g'_{ij} = e^{2\sigma} g_{ij}.
\end{equation}
Using the fact that $\mathcal{A}$ is Lifshitz invariant,
the relation \eq{Variation-W} can be integrated and we obtain the
transformation of the 
effective action under a finite Lifshitz scale transformation:
\begin{equation}\label{effective-action-prime}
    W(e^{z\sigma} N,e^{2\sigma} g_{ij};A)=W(N,g_{ij};A)+I_{\text{anom}}(A,
    N,g_{ij},\sigma),
\end{equation}
where
    \be
    I_{\text{anom}}(A, N,g_{ij},\sigma)=\int
    d^{d+1}x \,N\sqrt{g}\, \mathcal{I}(x)\,\sigma.
\ee
We are interested in the expectation value of the current in the presence
of a $U(1)$ gauge field background, which is given by
\begin{equation}
J^\mu(x) = \frac{1}{N\sqrt{g}}\,
\frac{\delta W}{\delta A_\mu(x)}.
\end{equation}
As a result, we obtain the transformation law for the current: 
\begin{equation}\label{current-prime}
J'^{\mu}(x)
= e^{-(z+d_s)\sigma} J^\mu (x)
+ \frac{1}{N'\sqrt{g'}}\,
  \frac{\delta}{\delta A_\mu (x)}\int d^{d+1}x\, N\sqrt{g}\,
  \mathcal{I}\,\sigma.
\end{equation}
Here the second term gives the anomalous current.
In the following
subsections, we
use this to determine the induced current
in a five dimensional BLFT due to an external
electromagnetic background.
As we will see, the derivation of
the anomaly induced current involves integrations by parts which
produce boundary terms. These terms vanish on closed manifolds but
become non zero in the presence of a boundary which gives rise to
surface currents on the boundary manifold. Similar to the discussion
in \cite{Chu:2018ksb} for $z=1$ case, the presence of these boundary
current is crucial as they are needed to maintain the global
conservation of charges. Moreover, the total current arising from the
bulk and the boundary current densities is finite through any interval
extended in the normal direction of the boundary as described later.
In the following, we will determine the bulk induced current followed by boundary current in a five dimensional BLFT.

\subsection{Anomaly induced current for 5D BLFT}\label{sec-5D-current}

We now turn our attention to the $(4+1)$-dimensional boundary Lifshitz
theory characterized by a dynamical critical exponent of $z=2$
(corresponding to $d_s=4$ spatial dimensions). Using the anisotropic
Lifshitz scale anomaly \eq{A-5D} for the 5D LFT, we can obtain from
\eq{current-prime} the transformation law of the current under the
scale transformation \eq{anisotropi-Wey-transf}. In the following, we
compute this separately for the temporal and spatial components.

 \subsubsection*{Anomalous current from the $c_S$ and the $c_T$ terms} 
Consider first the $c_T$ part of $I_{\rm anom}$ as
\begin{equation}\label{I_T}
I_{\rm anom} :=  c_T \int d^5x \, N\sqrt{g} \; \sigma\, F_{t i} F^{t i}.
\end{equation}
Using $\delta(F^2)=2F\delta F$ and $\delta F_{t i}=\nabla_t\delta A_i
-\nabla_i\delta A_t$,
and integrating by parts, we obtain the anomalous current
arising from the $c_T$-term:
\begin{equation}\label{ct-current}
J^{t}= -2  c_T\,\nabla_i(\sigma F^{i t}),\qquad
J^{i}= -2  c_T\,\nabla_t(\sigma F^{t i}).
\end{equation}
Next consider the spatial-derivative part
\begin{equation}
I_{\rm anom} :=  c_S \int d^5x \, N\sqrt{g} \; \sigma \, F_{ij}\,\nabla^2 F^{ij}.
\end{equation}
Variation of the above action gives
\begin{equation}
\delta I_{\rm anom}
=  c_S \int d^5x\, N\sqrt{g}\;\Big[\,\sigma\,\delta F_{ij}\,\nabla^2 F^{ij}
+ \sigma\,F_{ij}\,\nabla^2\delta F^{ij}\Big].
\end{equation}
On integrating the second term by parts twice so that the Laplacian
acts on $\sigma F_{ij}$ and using the product rule $\nabla^2(\sigma F)
= \sigma\nabla^2 F + 2(\nabla\sigma)(\nabla F) + (\nabla^2\sigma)F$,
one obtains the compact expression as
\begin{equation}\label{deltaIs}
\delta I_{\rm anom}
=  c_S \int d^5x\, N\sqrt{g}\;\delta F^{ij}\,
\big[\,2\sigma\,\nabla^2 F_{ij} + 2(\nabla^k\sigma)\nabla_k F_{ij} +
  (\nabla^2\sigma)\,F_{ij}\big].
\end{equation}
Here $\delta F^{ij}=2\nabla^{[i}\delta A^{j]}$.
Since the bracketed tensor is antisymmetric,
we can write
\begin{equation}
\delta I_{\rm anom} = 2  c_S \int d^5x\, N\sqrt{g}\; \nabla^{i}\delta A^{j}\; B_{ij},
\qquad
B_{ij} := 2\sigma\,\nabla^2 F_{ij} + 2(\nabla^k\sigma)\nabla_k F_{ij}
+ (\nabla^2\sigma)\,F_{ij}.
\end{equation}
Integrating $\nabla^{i}$ by parts yields
\begin{equation}
\delta I_{\rm anom} = -2  c_S \int d^5x\, N\sqrt{g}\; \delta A^{j} \,\nabla^{i}B_{ij}.
\end{equation}
Hence the $c_S$-term contributions to the current is given by
\begin{equation}\label{cs-variation-alt}
J^{j}
= -4  c_S\,\nabla_{i}\!\;\Big(\sigma\,\nabla^2 F^{\,i j}
+ (\nabla^k\sigma)\nabla_k F^{\,i j} + \tfrac{1}{2}(\nabla^2\sigma)\,
F^{\,i j}\Big).
\end{equation}
Note that the $c_S$ term does not contribute to the $J^t$ component.

Collecting the results \eqref{ct-current} and \eqref{cs-variation-alt}
we obtain the anomalous current:
\begin{align}
J^{t} &= -\;2  c_T\,\nabla_i\!\big(\sigma\,F^{i t}\big),\label{J-t}\\[6pt]
J^{i} &= -\;2  c_T\,\nabla_t\!\big(\sigma\,F^{t i}\big)\label{J-i}\\[4pt]
&\qquad\qquad -\,4  c_S\,\nabla_{j}\!\Big(\sigma\,\nabla^2 F^{j i}
+ (\nabla^k\sigma)\nabla_k F^{j i} + \tfrac{1}{2}(\nabla^2\sigma)F^{j i}\Big).
\nonumber
\end{align}
The first two lines are the direct $c_T$-anomaly analog of the
relativistic result and the last line is the contribution from the
higher-derivative spatial anomaly. This form of the current is fully
covariant with respect to the foliation geometry.

\subsubsection*{Anomalous boundary current}
  
Now to determine the current in BLFT on a flat spacetime, we utilize
the relation between a flat metric and a Lifshitz-like
geometry. Specifically, a BLFT defined on a flat half-space with a
planar boundary with the metric
\begin{equation}
    ds^2=-dt^2+dx^2+dy_a^2,\quad  x\ge 0, 
\end{equation}
can be obtained from the Lifshitz like metric
\begin{eqnarray}\label{CFTAdS}
ds^2=-\frac{dt^2}{x^{2z}}+\frac{dx^2+dy_a^2}{x^2},\ \  x\ge 0,
\end{eqnarray}
via the anisotropic rescaling \eq{anisotropi-Wey-transf} with
$\sigma=\ln x$. Here $x=0$ corresponds to the boundary of the half
space in the BLFT as well as the boundary of the Lifshitz geometry. By
utilizing this relation, we proceed to evaluate both the temporal and
spatial components of the anomalous current in the BLFT defined on
flat spacetime with a planar boundary.

By using the local anisotropic scale transformation with $\sigma = \ln x$ in
\eq{J-t}, we obtain the time component of current as follows
\begin{equation}
\begin{aligned}
J_{\mathrm{BLFT}}^{t}
&= - 2  c_T\, \partial_i\!\big( \sigma F^{i t} \big)
\\[4pt]
&=- 2  c_T \Big[ \frac{1}{x}\, F^{x t}
\;+\; (\ln x)\, \partial_i F^{i t} \Big].
\end{aligned}
\end{equation}
For small distance $x$ from the boundary,
the leading order  pure $F^{x t}$ term gives an inverse dependence on
$x$  similar to the case of BCFT as
\begin{equation}\label{t-pure}
J_{\mathrm{BLFT}}^{t}
= -2  c_T\,\frac{F^{x t}}{x}.
\end{equation}
Similarly, we obtain the full spatial component of current utilizing
$\sigma = \ln x$ in \eq{J-i} as
\begin{equation}\label{Current-spatial-5D}
\begin{aligned}
J_{\mathrm{BLFT}}^{i} &=  -2  c_T\,(\ln x)\,\partial_t F^{t i} \\
&\quad -4  c_S \Big[
(\ln x)\,\partial_j\partial^2 F^{j i}
+ \frac{1}{x}\,\partial^2 F^{x i}
+ \frac{1}{x}\,\partial_j\partial_x F^{j i}
- \frac{1}{x^2}\,\partial_x F^{x i}
- \frac{1}{2x^2}\,\partial_j F^{j i}
+ \frac{1}{x^3}\,F^{x i}
\Big].
\end{aligned}
\end{equation}
We see that all terms are linear in $F$ and its derivatives up to
third order.  In particular, the pure $F^{x y}$ (no derivatives acting
on $F$) term appears only in the transverse component (i.e. $i=y$) in
the above expression, so at leading order
\begin{equation}\label{i-pure}
J_{\mathrm{BLFT}}^{y} = -4 c_S\,\frac{F^{x y}}{x^3}.
\end{equation}

Our results show that the temporal and spatial components of the
boundary current
exhibit distinct scaling behaviors with respect to $x$. This behavior
is expected as time and space scale differently in Lifshitz field
theories. They are also determined from the central charge of the
Lifshitz scale anomaly which is itself related to the beta function of
the theory.
The contributions involving $F$ and its derivatives
represent universal features of the induced current in five
dimensional BLFTs. 
Note that the dependence of the coefficients $c_{T,S}$ on boundary conditions is not known a priori and determining this requires an explicit computation within a specific model. The holographic model studied in the following section provides an example of how this dependence appears in a BLFT.

\subsection{Finiteness of total current}

We now compute the boundary current and show that the total current
is finite for BLFT similar to the case of BCFT \cite{Chu:2018ksb}. We observed that the near boundary bulk currents given in \eq{t-pure} and \eq{i-pure} diverge as $x\to 0$. These divergences are regularised by a cutoff $x = \epsilon$ and cancelled by surface currents $j^\mu$ localised at the boundary. Due to gauge invariance, the conservation laws of currents can be written as
\begin{align}
    &\text{Bulk:} \quad \partial_t J^t + \partial_x J^x + \partial_a J^a = 0,
    \label{bulk-current}\\
    &\text{Boundary:} \quad \partial_t j^t + \partial_a j^a = -J^x\big|_{x=\epsilon},
    \label{bdy-current}
\end{align}
where $J^x$ is the normal component of the bulk current.

\paragraph{Temporal component:} On varying the $c_T$ part of $I_{\rm anom}$ from \eq{I_T} with respect to $A_t$ and integrating by parts with inward-pointing normal $n_x = 1$ at $x = \epsilon$, we obtain
\begin{equation}
    \delta I_{\mathrm{anom}}
    = 2c_T\int_\epsilon^\infty\!\partial_i(\sigma F^{ti})\,\delta A_t\,d^5x
    + 2c_T\int_{x=\epsilon}\!\!F^{tx}\sigma\,\delta A_t\,d^4x_\perp.
\end{equation}
The boundary term identifies the surface current at the cutoff as
\begin{equation}
    j^t = -2c_T\,F^{xt}\ln\epsilon.
\end{equation}
This diverges logarithmically similar to the BCFT case ($z=1$).
Now integrating the total current density $J^t_{\rm total} = J^t_{\rm bulk} +
\delta(x-\epsilon)\,j^t$ over the normal direction from $\epsilon$ to some depth $L$ gives
\begin{equation}
    \int_\epsilon^L J^t_{\rm total}\,dx
    = -2c_T F^{xt}\ln\!\frac{L}{\epsilon} + \bigl(-2c_T F^{xt}\ln\epsilon\bigr)
    = -2c_T F^{xt}\ln L,
\end{equation}
where the logarithmic divergences cancel exactly yielding a finite result.

\paragraph{Spatial component:} The spatial component exhibits a power-law divergence ($1/x^3$ instead of $1/x$), a qualitatively new feature arising from the Lifshitz anisotropy $z=2$. Now taking $i=x$ in \eq{Current-spatial-5D}, the leading singular term of the normal current is
\begin{equation}
    J^x\big|_{x=\epsilon} = \frac{2c_S\,\partial_a F^{ax}}{\epsilon^2} + \mathcal{O}\left(\tfrac{1}{\epsilon}\right).
\end{equation}
For a static background, the boundary conservation law \eq{bdy-current} gives
\begin{equation}
    j^y = \frac{2c_S\, F^{xy}}{\epsilon^2}.
\end{equation}
Then the full spatial current density is given by
\begin{equation}
    J^y_{\rm total}
    = -\frac{4c_S\,F^{xy}}{x^3}
    + \delta(x-\epsilon)\cdot\frac{2c_S\,F^{xy}}{\epsilon^2}
    + \mathcal{O}\left(\frac{1}{x^2}\right).
\end{equation}
On integrating the total spatial current density over the normal direction then gives
\begin{align}
    \int_\epsilon^L J^y_{\rm total}\,dx= -4c_S F^{xy}\left(-\frac{1}{2L^2}+\frac{1}{2\epsilon^2}\right)
    + \frac{2c_S F^{xy}}{\epsilon^2}
    = \frac{2c_S F^{xy}}{L^2},
\end{align}
where the leading $1/\epsilon^2$ divergences cancel exactly and total current remain finite.

\section{Holographic current for boundary Lifshitz field theory}
\label{sec-Hol-currrent}

Following our field theory analysis, we now employ holographic methods
to obtain the current for holographic BLFT in general dimensions.

\subsection{Review of holographic BLFT}

As shown in \cite{Chu:2024nwf}, the holographic bulk dual to Lifshitz
field theories defined on a $d$-dimensional manifold $M$ with a
boundary $\del M$ is described by the gravity theory living in a
portion of the $(d+1)$-dimensional Lifshitz spacetime $N$ bounded by
an end of the world (EOW) brane $Q$ such that $\del N = M \cup Q$. The
bulk action corresponding to this holographic dual is given by
\be\label{BLFT-action}
\begin{aligned}
  I_1=& \frac{1}{16\pi G_N}\int_N d^{d+1}x\sqrt{-g}
  \left(R - 2\L -\frac{\mathcal{F}_m^2}{4}
  -\frac{1}{2}M^2 \mathcal{A}_m^2\right)+\frac{1}{8\pi G_N}
  \int_Q d^{d}x \sqrt{-h}(K-T) \\
  &+\frac{1}{16\pi G_N}\int_Q \sqrt{-h} \;  \frac{1}{2}\mu M^2 a_m^2
  \, \cos \th ,
\end{aligned}
\ee
where $K$ is the scalar extrinsic curvature of $Q$, $h_{ij}$ is the
induced metric on $Q$ and $T$ is the tension of EOW brane $Q$. Since
the holographic dual of a Lifshitz field theory (LFT) incorporates a
massive vector field $\mathcal{A}_m$, the action must be supplemented
with a boundary term to ensure a well defined variation
principle. This term involves the projected gauge field $a_\mu$ on the
EOW brane $Q$. Within this framework, $\theta$ represents the angle
between the normal vectors of the brane $Q$ and the
manifold $M$, while $\mu$ is a constant defined as
\be
\mu = \frac{1}{d-1} \sqrt{\frac{f(z)}{2 |\L|}},
\ee
where $ f(z) = z^2 + (d-2)z + (d-1)^2$.
The metric  and the gauge field 
\begin{equation}
  ds^2=L^2\left[-\frac{dt^2}{r^{2z}}+\frac{dr^2}{r^2}
    +\frac{d \vec{x}^2}{r^2}\right], \qquad
  A = \sqrt{\frac{2(z-1)}{z}}\frac{L}{r^z}dt
\end{equation}
solves the equation of motion if the mass is fixed to be a specific value
\be
M^2=\frac{(d-1)z}{L^2},
\ee
and $L$ is related to the cosmological constant as
\be \label{Lambda}
\L = - \frac{z^2+(d-2)z+(d-1)^2}{2L^2}: = -\frac{f(z)}{2 L^2}.
\ee

Next consider the EOW brane $Q$. 
The variation of
\eq{BLFT-action} gives rise to the different boundary conditions such
as Neumann boundary condition (NBC) and conformal boundary condition
(CBC) on $Q$:
\bea
{\rm NBC:} && K_{\a\b}-(K-T)h_{\a\b}=T^A_{\a\b}, \label{nbc}\\
{\rm CBC:}  && d \,T-(d-1)K=T^A, \label{cbc}
\eea
where
\begin{equation}\label{TA}
  T^A_{\a\b}=\frac{1}{2}\mu M^2 \cos \th
  \left(a_\a a_\b-\frac{1}{2}h_{\a\b} a^2\right),
  \quad T^A=\left(1-\frac{d}{2}\right) \frac{1}{2}\mu M^2 \cos \th \; a^2,
\end{equation}
is the boundary energy-momentum tensor coming from the massive gauge
vector on the EOW brane. Similarly, the boundary conditions for the
massive vector field can be obtained as
\begin{equation}\label{m-A-bc}
n_Q^N F_{N \a} + \mu M^2 \cos \th \,  a_\a \Big|_Q=0.
\end{equation}

Let us now consider a $d$-dimensional BLFT defined on a half space
$x\geq 0$. The bulk geometry is then given by a portion of
$(d+1)$-dimensional Lifshitz spacetime as follows
\begin{equation}
  ds^2=L^2\left[-\frac{dt^2}{r^{2z}}+\frac{dr^2}{r^2}
    +\frac{dx^2}{r^2}+\frac{d \vec{y}^2}{r^2}\right],
\end{equation}
with coordinates $(t,r,x,y^i)$ and $i=1,...,d-2$.
One can show that the  EOW brane profile
\be \label{brane-profile}
x = r \sinh \left(\frac{\rho_*}{L}\right),
\ee
solves both the CBC \eq{cbc} and the NBC \eq{nbc},
where here
$\rho_*$ is related to the brane tension $T$ by
\be \label{tension-CBC}
T =\frac{z+2d-3}{2 L}\tanh \left(\frac{\rho_*}{L}\right).
\ee
At the same time, the non-vanishing component of vector field $a_\a$ 
\begin{equation} \label{a-component}
a_0=\sqrt{\frac{2(z-1)}{z}}\frac{L}{r^z},
\end{equation}
solves also the boundary condition \eq{m-A-bc}.
Note that the wedge 
solution
\eq{brane-profile} was originally obtained for the AdS/BCFT
\cite{Takayanagi:2011zk,Fujita:2011fp}. It is remarkable that the same
solution works
also for the case of holographic BLFT dual \cite{Chu:2024nwf} when a massive gauge field
is in presence.
Note also that the solutions \eq{brane-profile} and
\eq{tension-CBC} was originally
obtained in \cite{Chu:2024nwf} for
holographic dual of
BLFT${}_2$. It is interesting that 
the same wedge solution also works for the higher dimensions,
particularly since the NBC is characterized by a system
of multiple equations giving multiple constraints whereas the CBC is
governed by a single equation having only single constraint.
This happens because in higher dimensions, the additional
$y^i$-components of \eq{nbc}
leads to a single equation which coincide with what one already obtained
in \eq{cbc}.

\subsection{Holographic current}

We now extend the holographic BLFT model in \eq{BLFT-action} by
introducing a $U(1)$ gauge field in order to study the renormalized
current in these theories. The contribution of the gauge field to the
bulk action is given by

\begin{equation}
    I_2=\frac{1}{16\pi G_N}\int_N d^{d+1}x\sqrt{-g}
  \,\mathcal{F}_{\mu\nu}\mathcal{F}^{\mu\nu},
\end{equation}
where
$\mathcal{F}_{\mu\nu}=\partial_\mu\mathcal{A}_\nu-\partial_\nu\mathcal{A}_\mu$
denotes the field strength of the massless $U(1)$ gauge field
$\mathcal{A}_\mu$. Consequently, the total action of the system is
$I=I_1+I_2$. For a massless gauge field, the action $I_2$ is invariant
under $U(1)$ gauge transformations in the bulk. However, to ensure
that the variation principle is well defined in the presence of
boundaries, one must impose appropriate boundary conditions on the
gauge field. Varying the action with respect to $\mathcal{A}_\mu$
produces a boundary term which leads to the boundary condition for the
massless gauge field. The boundary conditions on $Q$ read
\begin{equation}\label{bdy-condition-massless}
\mathcal{F}_{\mu\nu}n_Q^{\mu}\Pi^{\nu}_{\ \alpha}=0,
\end{equation}
where $n_Q$ denotes the inward-pointing normal vector on $Q$ and $\Pi$
represents the projection operator that gives the vector field and
metric on $Q$:
$\bar{A}_{\alpha}=\Pi^{\nu}_{\ \alpha}\mathcal{A}_{\mu}$ and
$\gamma_{\alpha\beta}=\Pi^{\mu}_{\ \alpha}\Pi^{\nu}_{\ \beta}G_{\mu\nu}$.
Given the planar symmetry of the boundary, we restrict our attention
to gauge field configurations $\mathcal{A}_\mu$ that depend only on
the coordinates $r$ and $x$.
We therefore choose the following ansatz for the massless gauge field as
\begin{equation}
\mathcal A_r=\mathcal A_r(r),\qquad
\mathcal A_x=\mathcal A_x(x),\qquad
\mathcal A_t=\mathcal A_t(r,x),\qquad
\mathcal A_{y^i}=\mathcal A_{y^i}(r,x),
\end{equation}
and assume no dependence on $t$ or $y^i$ (so $\partial_t=\partial_{y^i}=0$).
The source free Maxwell equations then take the form
\begin{equation}
  \nabla_\mu\mathcal F^{\mu\nu}=\frac{1}{\sqrt g}\,\partial_\mu\!\big(\sqrt g\,
  \mathcal F^{\mu\nu}\big)=0.
\end{equation}
The inverse metric components and the volume factor are given by
\begin{equation}
g^{tt}=\frac{r^{2z}}{L^2},\qquad g^{rr}=g^{xx}=g^{y^i y^i}=\frac{r^2}{L^2},
\qquad
\sqrt{g}=L^{\,d+1}r^{-(z+d)}.
\end{equation}
Since there is anisotropy in time and spatial directions, the Maxwell
equations decompose into the following independent components:

\noindent For $\nu=t$:
\begin{equation}\label{A_t-eqn}
  r\partial_r^2\mathcal A_t -(d-z-2)\,\partial_r\mathcal A_t +
  r\partial_x^2\mathcal A_t = 0.
\end{equation}

\noindent For $\nu=y^i$ (each transverse index):

\begin{equation}\label{A_i-eqn}
  r\partial_r^2\mathcal A_{y^i} -(d+z-4)\,\partial_r\mathcal A_{y^i} +
  r\partial_x^2\mathcal A_{y^i} = 0,
\qquad i=1,\dots,d-2.
\end{equation}

\noindent For $\nu=r, x$: all potentially contributing terms vanishes
identically.
So we see that the only non-trivial Maxwell PDEs are those for
$\mathcal A_t$ and each $\mathcal A_{y^i}$ as given above.  Similarly,
the boundary condition \eq{bdy-condition-massless} for the gauge field
$\mathcal{A}_\mu$ can be obtained as
\begin{equation}\label{bdy-gauge-field}
 ( \del_x \mathcal{A}_a-\sinh\frac{\rho_*}{L}\,
  \del_r \mathcal{A}_a ) \Big|_{x=r \sinh \frac{\rho_*}{L} } =0, \qquad a=t,y^i.
\end{equation}

In order to solve \eq{A_t-eqn} and \eq{A_i-eqn}, we take the ansatz for the
vector field analogous to that used in \cite{Miao:2017aba,Chu:2018ntx} as
follows
\begin{equation}\label{vector-ansatz}
  \mathcal{A}_a=A^{(0)}_a+  x f_1(\frac{r}{x}) A^{(1)}_a
  + x^2 f_2(\frac{r}{x}) A^{(2)}_a+
  \cdots,
\end{equation}
where the condition $f_i (0)=1$ ensures that $\mathcal{A}_a$ matches
the boundary value $A_a$ at $r=0$, with $A^{(i)}$ acting as the
coefficients of the near boundary expansion:
\begin{eqnarray}\label{vectorgauge}
A_a=A^{(0)}_a+ xA^{(1)}_a+ \cdots .
\end{eqnarray}
The coefficient $A^{(1)}_a$ is fixed by the field strength evaluated
at the boundary as
\begin{equation}
    A^{(1)}_a=F_{xa}=F_{na}.
\end{equation}
With \eq{vector-ansatz}, the Maxwell equation \eq{A_t-eqn} and the
boundary condition \eq{bdy-gauge-field} read
\begin{align} \label{ode1}
    s(s^2+1)f''(s)-(d-z-2)f'(s)&=0, \quad s=r/x,
\end{align}
\begin{equation}\label{bdy-Q-f(s)}
  \cosh\frac{\rho_*}{L}\coth\frac{\rho_*}{L}\,
  f'(\mathrm{csch} \frac{\rho_*}{L})=f(\mathrm{csch} \frac{\rho_*}{L}).
\end{equation}
The ODE \eq{ode1} has the general solution
\begin{equation}
  f(s)=c_1+\frac{s^{d-z-1}\,c_2\, _2F_1\left(\frac{d-z-2}{2},
    \frac{d-z-1}{2}; \frac{d-z+1}{2};-s^2\right)}{d-z-1},
\end{equation}
where $c_1$ and $c_2$ are integration constants.
The condition $f(0)=1$ gives $c_1=1$ and so
\begin{equation}\label{f-At}
  f(s)=1+\frac{s^{d-z-1}\,\alpha_d^T\, _2F_1\left(\frac{d-z-2}{2},
    \frac{d-z-1}{2}; \frac{d-z+1}{2};-s^2\right)}{d-z-1}.
\end{equation}
The boundary condition \eq{bdy-Q-f(s)}
allows $\alpha_d^T$ to be solved and one finds
\begin{equation}
  \alpha_d^T= \frac{(-d+z+1) \coth\left(\frac{\rho_*}{L}\right)^{d}
    \text{csch}^{-d+z+1}\left(\frac{\rho_* }{L}\right)}{\coth
    \left(\frac{\rho_*}{L}\right)^{d} \,
    _2F_1\left(\frac{d-z-2}{2},\frac{d-z-1}{2};\frac{d-z+1}{2};
    -\text{csch}^2(\frac{\rho_*}{L})\right)+(-d+z+1)\cosh
    ^2\left(\frac{\rho_*}{L}\right)
    \coth\left(\frac{\rho_*}{L}\right)^{z+2}}.
\end{equation}
Similarly, with the ansatz \eq{vector-ansatz}, the 
Maxwell equation \eq{A_i-eqn} for $\mathcal{A}_i$ reads
\begin{equation}\label{f-Ai}
    s(s^2+1)f''(s)-(d+z-4)f'(s)=0.
\end{equation}
This has the general solution 
\begin{equation}\label{f-Ai-1}
  f(s)=1+\frac{s^{d+z-3}\,\alpha_d^S\, _2F_1\left(\frac{d+z-4}{2},
    \frac{d+z-3}{2};\frac{d+z-1}{2};-s^2\right)}{d+z-3}
\end{equation}
and the integration constant $\alpha_d^S$ can be solved using 
\eq{bdy-Q-f(s)}:
\begin{equation}
    \alpha_d^S=\frac{-(d+z-3) \coth\left(\frac{\rho_*}{L}\right)^{d}
      \text{csch}^{-d-z+3}\left(\frac{\rho_*}{L}\right)}
          {\coth\left(\frac{\rho_*}{L}\right)^{d}
      \,
      _2F_1\left(\frac{d+z-4}{2},\frac{d+z-3}{2};
      \frac{d+z-1}{2};-\text{csch}^2\left(\frac{\rho_*}{L}\right)\right)
      + (-d-z+3)
      \cosh ^2\left(\frac{\rho_*}{L}\right) \coth
      ^{4-z}\left(\frac{\rho_*}{L}\right)}.
\end{equation}
We note that for $z=1$, the integration constants
agrees: $\alpha_d^T= \alpha_d^S$  for arbitrary $d$.
In particular, they reduce to the holographic BCFT result \cite{Chu:2018ntx} 
$\alpha_4^T= \alpha_4^S=1$ for $d=4$.

The holographic current can be obtained from the holographic model of
BLFT described by the action $I$ as follows
\begin{equation}\label{hol-current-def}
  \la J^a \ra=\lim_{r\to 0}\frac{\delta I}{\delta A_a}
  =\lim_{r\to 0}\sqrt{G}\mathcal{F}^{ra}.
\end{equation}
Now using \eq{f-At}, \eq{f-Ai-1} and \eq{vector-ansatz} in the above
expression for current \eq{hol-current-def}, the temporal and spatial
component of the current in the leading order can be obtained as
    \begin{align}
        \la J_t \ra
        &=-\alpha_d^T\frac{F_{tn}}{x^{d-z-2}}+\mathcal{O}(\frac{1}{x^{d-z-3}})
        \label{C-Jt},\\ \la
        J_i \ra
        &=-\alpha_d^S\frac{F_{in}}{x^{d+z-4}}+\mathcal{O}(\frac{1}{x^{d+z-5}})
        \label{C-Ji}.
    \end{align}
We observe that the inherent anisotropy of BLFTs manifests as distinct
scaling behaviors for the current components relative to the distance
from the boundary $x$. This contrasts with holographic BCFTs where
$z=1$ and the current displays a uniform dependence on $x$ across all
components \cite{Chu:2018ntx}.

In the appendix \ref{dim}, we show that if the anisotropy $z$ is related to 
the spatial dimension as
$z= d_s - 2$, the dual field theory is critical and
exhibits scale invariance. It is interesting to note that in this case
$\alpha_{d}^T=1$, meaning that 
$\alpha_d^T$ is unaffected by the boundary
conditions, while $\a_d^S$  maintains a dependence on $\rho_*$.
We also note that the holographic current takes the following form
\begin{equation}\label{crit-dim}
    \begin{aligned}
      \la J_t \ra =-\frac{F_{tn}}{x}+ \cdots,
      \quad\la J_i \ra =-\alpha_{d}^S\frac{F_{in}}{x^{2z-1}}+ \cdots \,\,\,\,.
    \end{aligned}
\end{equation}
In particular the leading order $1/x$ behavior of $J_t$ is universal and
is independent of $d$.

\subsection{Holographic current for dual 5D BLFT}

As discussed in the earlier section where we derived the current for
5D boundary Lifshitz field theory with $z=2$. The corresponding
holographic bulk dual is now $(5+1)$-dimensional with $z=2$.
In this case,  the integration constants are given by
\begin{equation}\label{constant-5D}
  \alpha_5^T=1 , \quad\alpha_5^S=\frac{1}{2}\frac{1}
        { 1-\tanh\left(\frac{\rho_* }{L}\right) }.
\end{equation}
The holographic current can now be obtained from \eq{crit-dim} as
\begin{equation}
    \begin{aligned}
        \la J_t \ra &=-\frac{F_{tn}}{x}+\mathcal{O}(1),\\
        \la J_i \ra &=-\alpha_5^S\frac{F_{in}}{x^{3}}+\mathcal{O}
        (\frac{1}{x^{2}}),
    \end{aligned}
\end{equation}
where $\alpha_5^S$ is given in \eq{constant-5D}.
As noted, the
temporal component of the near boundary current is independent of the
choice of boundary conditions. Interestingly, the holographic current
reproduces the same scaling with $x$ from the dual field theory side
given in \eq{t-pure} and \eq{i-pure}. Furthermore,
we can identify the central charges
\be
c_T = -1/2, \quad c_S=-\alpha_5^S/4
\ee
for the anisotropic Lifshitz scale anomaly.

\section{Summary and discussion}\label{sec-summary}

To summarize, we have studied the anisotropic Lifshitz scale anomalies
and their physical consequences in boundary Lifshitz field theories
(BLFTs) in the presence of an external $U(1)$ gauge field. Unlike
relativistic quantum field theories, where scale invariance is
generically enhanced to full conformal symmetry, non-relativistic
theories with Lifshitz scaling possess intrinsically anisotropic
symmetries characterized by a dynamical exponent. This anisotropy
leads to novel anomaly structures and boundary phenomena which are
different from their relativistic counterparts. We focused on Lifshitz
field theories with anisotropic Lifshitz scale symmetry and analyzed
how quantum violations of this symmetry manifest near boundaries. In
this context, we showed that the anisotropic Lifshitz scale anomaly
induces an anomalous near boundary current in BLFTs. Our approach was
based on the transformation law of current under the anisotropic scale
transformations.

We first performed a systematic weight analysis to identify the
admissible anomaly structures and then computed the induced current in
a BLFT defined on flat spacetime with a planar boundary. Exploiting
the relation between flat space and Lifshitz-like spacetime via
anisotropic rescaling, we explicitly determined both the temporal and
spatial components of the current. A key result is that these
components exhibit different power law dependence on the distance from
the boundary which shows the inherent anisotropy of space and time in
BLFTs. This stands in stark contrast to boundary QFTs where the
relativistic symmetry ensures that both temporal and spatial
components share an identical dependence. Furthermore, the dependence
of the current on the central charge of the anisotropic Lifshitz scale
anomaly reflects its connection to the beta function of the theory.

Subsequently, we obtained the current holographically from the bulk
dual of BLFTs. By extending the gravitational action to include a
Maxwell-type $U(1)$ gauge sector, the current is derived from the bulk
dynamics of a massless gauge field. Solving the source free Maxwell
equations in the bulk and imposing suitable boundary conditions on the
EOW brane, we evaluated the holographic current in the vicinity of the
boundary.  This holographic analysis showed that in dual 5D BLFTs, the
temporal component of the current is insensitive to boundary
conditions while the spatial component depends explicitly on
them. Furthermore, the temporal sector allows us to extract the
central charges associated with the anisotropic Lifshitz scale anomaly
establishing a direct correspondence between the induced current and
the anomaly coefficient. Remarkably, we observed that the holographic
computation reproduces the same leading order dependence on the
distance from the boundary as found in the dual field theory result
providing a consistency check of our analysis.

We further observed that the current becomes singular as one approaches the boundary which can be understood as a UV/regulator effect: the anomaly induced response is localized within a boundary layer whose thickness vanishes in the continuum limit. With a finite microscopic cutoff (e.g. lattice spacing), the current remains large but finite and concentrated near the surface while physically meaningful quantities such as the integrated edge current remain finite and may contain universal, anomaly determined contributions. This singular behavior indicates that the current is not a bulk property but is instead localized near the boundary analogous to the Casimir effect where boundary conditions restrict vacuum modes. In the relativistic ($z=1$) case, boundary contributions cancel the apparent divergence of the bulk current at $x=0$ leading to a finite total current \cite{Chu:2018ksb}. A similar situation arises in the Lifshitz case ($z \neq 1$) where anisotropic scaling modifies the power law behavior of the divergence without changing the underlying physical mechanism.

Overall, this study establishes a direct link between anisotropic
Lifshitz scale anomalies and boundary induced currents in Lifshitz
field theories, both from field theory and holographic
perspectives. Our results highlight qualitative differences from
relativistic boundary conformal field theories (BCFTs) and provide new
insight into universal boundary effects in non-relativistic quantum
systems. There are several interesting directions to follow from the
present work. 
It would be interesting to derive the Lifshitz scale anomaly through holographic renormalization along the lines of \cite{Griffin:2011xs,Baggio:2011ha,Baggio:2011cp}.
Another interesting direction would be to explore the role
of other external background fields such as curved spatial metrics,
torsion, or Newton-Cartan geometries
\cite{Auzzi:2015fgg,Arav:2016xjc}. Coupling LFTs to such backgrounds
may lead to additional anomaly induced responses including energy or
momentum currents, thereby deepening our understanding of
non-relativistic anomaly induced transport phenomena.  The present
work focuses on planar boundaries however extending the analysis to
curved or dynamical boundaries could reveal new boundary localized
effects and geometric contributions to the induced current. In
particular, understanding how extrinsic curvature terms enter the
anomaly and influence transport remains an open question. It would
also be interesting to derive the anomalous current using alternative
approaches such as the Green’s function method which may provide
complementary insights.  Another important direction is to explore
more complex boundary geometries or non-planar EOW branes in the
holographic setup which may uncover new universality classes of
boundary responses.  Finally, it would be highly interesting to
investigate potential connections with condensed matter systems
exhibiting Lifshitz scaling such as quantum critical points and to
assess whether the boundary currents identified in this work can lead
to observable experimental signatures. We leave these interesting
questions for future investigation.

\section*{Acknowledgments}
C.S.C acknowledge the support of this work from NCTS, the
National Science and Technology Council of Taiwan for the
grant 113-2112-M-007-039-MY3,
and the National Tsing Hua University 2025 Talent
Development Fund for the TSAI WANG, YUAN-YANG Distinguished Talent
Chair professorship.
H.P acknowledges the support of this work by NCTS. H.P. was supported by the National Research Foundation of Korea (NRF) grant funded by the Korean government (MSIT) (RS-2022-NR069038). H.P. was supported by Brain Pool program funded by the Ministry of Science and ICT through the National Research Foundation of Korea (RS-2025-25457100).

\begin{appendices}
  \section{Anisotropic weight of the gauge field}\label{dim}
  Consider Lifshitz field theory in $(d_s+1)$-dimensions where
  $d_s$ is the number
  of spatial dimension.
We define the scaling weights in terms of momentum (mass) dimensions $[M]$. Under the transformation
\begin{equation}
t\mapsto \lambda^{z}t,\qquad x^{i}\mapsto \lambda\,x^{i},
\end{equation}
the dimension of coordinates are $ [t] = -z,[x] = -1$ and consequently for the
derivatives:  $[\partial_{t}]=z$ and $ [\partial_{i}]=1$ \cite{Horava:2009uw, Horava:2008jf}.
From minimal coupling we require the covariant derivatives
\begin{equation}
D_{t}=\partial_{t}-i  A_{t},\qquad D_{i}=\partial_{i}-i  A_{i}
\end{equation}
to scale like the corresponding derivatives.  Hence we have
\begin{equation}
[A_{t}]=[\partial_{t}]=z,\qquad [A_{i}]=[\partial_{i}]=1.
\end{equation}
Consider the following gauge kinetic term 
\begin{equation}
S_{\rm gauge} =-\frac{1}{4e^2}\int dt\,d^{d_s}x\; F_{t i}F^{t i},\qquad 
F_{t i}=\partial_{t}A_{i}-\partial_{i}A_{t},
\end{equation}
then
\begin{equation}
[F_{t i}] = [\partial_{t}] + [A_{i}] = z + 1.
\end{equation}
or equivalently $[\partial_{i}]+[A_{t}]=1+z$, and the two expressions
agree. Upon requiring the Lagrangian density to have weight equal to
the measure weight $z+d_s$ yields
\begin{equation}
2\big(z+1-[e]\big) = z + d_s.
\end{equation}
Solving for $[e]$ gives
\begin{equation}\label{e-dim}
\qquad [e] = \dfrac{z+2-d_s}{2}. 
\end{equation}
Also, the weight of spatial component of the field strength is given by
\begin{equation}
[F_{ij}] = [\partial_{i}] + [A_{j}] = 2.
\end{equation}
From the above analysis, we observe the following:

\begin{itemize}
  \item The \emph{critical} spatial dimension where the gauge coupling
    is dimensionless (marginal) is
\begin{equation}\label{d-crit}
  d_s= z + 2,
\end{equation}
  since then $[e]=0$.  This reproduces the relativistic case $z=1$,
  $d_s=3$ (i.e.\ $3+1$ spacetime). Note that the relation
  \eq{d-crit} is different from the generalized Lifshitz model
  introduced in \cite{Keranen:2016ija} where $d_s=z$.
  \item If $[e]>0$ the coupling is relevant (super-renormalizable) and
    typically one does not expect universal logarithmic running from
    UV loops (no marginal log). If $[e]<0$ the coupling is irrelevant
    (non-renormalizable).
\end{itemize}

In the following, we show examples of weights for a few cases of interest:

\begin{itemize}
  \item CFT and BCFT in $(3+1)$-dimensions ($z=1, d_s=3$):
\begin{equation}
  [e]=0,\qquad [A_t]=[A_i]=1,\qquad [F_{ti}]=[F_{ij}]=2.
\end{equation}

\item Lifshitz theory in $(3+1)$-dimensions ($z=2, d_s=3$):
\begin{equation}
  [e]=\frac{1}{2}, \qquad [A_t]=2\qquad [A_i]=1,\qquad [F_{ti}]=3,\qquad
  [F_{ij}]=2.
  \end{equation}
  
  \item QLM in $(2+1)$-dimensions ($z=2, d_s=2$):
\begin{equation}
  [e]=1, \qquad [A_t]=2, \qquad [A_i]=1,\qquad [F_{ti}]=3,\qquad [F_{ij}]=2.
\end{equation}

  \item Critical LFT or BLFT in $(4+1)$-dimensions ($z=2, d_s=4$):
\begin{equation}
  [e]=0, \qquad [A_t]=2,\qquad [A_i]=1, \qquad [F_{ti}]=3,\qquad [F_{ij}]=2.
\end{equation}
\end{itemize}
We have considered the last case for our analysis in \cref{sec-5D-current}.

  \section{Cohomological analysis}\label{Coh-analysis}
  
We now solve the cohomology problem explicitly in the case of a flat background
with a $U(1)$ gauge field, showing that the two terms in \eq{A-5D} are
the complete set of non-trivial anomalies. Working with the BRST formalism similar to \cite{Griffin:2011xs} (see also \cite{Baggio:2011ha,Arav:2014goa}) the
scaling ghost $c$ (constant for global Lifshitz scaling) and gauge ghost
$\lambda(x)$ are introduced, with the total BRST operator $s = s_\sigma +
s_\lambda$ acting as 
\begin{equation}
sA_t = zcA_t + \partial_t\lambda, \quad sA_i = cA_i +\partial_i\lambda, \quad sc = s\lambda = 0. 
\end{equation}
Since we restrict ourselves to a flat background and consider only global Lifshitz scaling as discussed in \cref{sec-An-Weyl}, the global anisotropic (Weyl) transformation is realized as the corresponding global Lifshitz coordinate dilatation. Consequently, the BRST operator $s_\sigma$ acts not only on the fields but also on coordinate derivatives and composite expressions, by multiplying by $c$ times their total anisotropic weight. In particular, $s_\sigma(\partial_t\lambda) = zc\,\partial_t\lambda$ since
$[\partial_t\lambda] = z$, even though $s_\sigma\lambda = 0$ since $[\lambda] =0$.
Therefore, using the graded
Leibniz rule (since $c$ is Grassmann-odd, so $sc = 0$ and $c^2=0$):
\begin{align}
    s^2A_t
    &= s(zcA_t + \partial_t\lambda) \notag\\
    &= -zc\,(zcA_t + \partial_t\lambda)
      + s_\sigma(\partial_t\lambda)
      + s_\lambda(\partial_t\lambda)
    \notag\\
    &= -z^2c^2A_t
      - zc\,\partial_t\lambda
      + zc\,\partial_t\lambda
    \;=\; 0. 
\end{align}
and likewise $s^2A_i=0$. The BRST operator is therefore nilpotent i.e. $s^2=0$.

A candidate anomaly at ghost number $+1$
takes the form $\mathcal{A} = c\int d^{d+1}x\,\mathcal{I}_0[F,\partial F,\ldots]$,
where gauge invariance requires $\mathcal{I}_0$ to be built from the field
strengths $F_{ti}$, $F_{ij}$ and their spatial derivatives, and Lifshitz
invariance requires $\mathcal{I}_0$ to carry anisotropic weight exactly $(z+d_s)$,
so that the integrated density is marginal. The cocycle condition $s\mathcal{A} =
0$ is then automatically satisfied. The cohomology problem reduces to identifying
which such $\mathcal{A}$ are not of the form $s\mathcal{B}$ for a
ghost number zero local functional $\mathcal{B}$. A coboundary $s\mathcal{B} =
c\,\delta_\sigma\mathcal{B}$ requires a gauge-invariant local functional
$\mathcal{B} = \int \mathcal{L}$ satisfying $\delta_\sigma \mathcal{B} = \int
\mathcal{I}_0$. However, for any marginal gauge-invariant counterterm $\mathcal{B}$
(built from $F_{\mu\nu}$ and derivatives, with $\mathcal{L}$ of weight $z+d_s$),
the global scaling variation yields
\begin{equation}
    \delta_\sigma \mathcal{B}
    = \sigma \int \bigl(w - (z+d_s)\bigr)\mathcal{L}
    + \text{boundary terms} = 0,
\end{equation}
since marginal operators are scale-invariant. Sub-marginal counterterms (weight:
$w< z+d_s$) produce variations of weight strictly less than $(z+d_s)$ which cannot
reproduce a weight-$(z+d_s)$ density. Hence no gauge-invariant local
counterterm can cancel a marginal anomaly term: every gauge-invariant,
Lifshitz scale invariant density at weight $(z+d_s)$ modulo total spatial
derivatives is a genuine non-trivial cohomology element. For $(d_s,z) = (4,2)$,
the marginal weight is $z + d_s = 6$. A complete enumeration of gauge-invariant
densities at this weight, built from $F_{ti}$ (weight $3$) and $F_{ij}$
(weight $2$) with spatial derivatives $\partial_i$ (weight $1$) and at most
quadratic order in $F$, yields:
\begin{enumerate}
    \item $F_{ti}F^{ti}$ with weight $3 + 3 = 6$.
    \item $F_{ij}\nabla^2 F^{ij}$ with weight $2 + 2 + 2 = 6$.
    \item The cubic candidate $F_{ij}F^{ik}F_{k}{}^{j}$, which vanishes
          identically by antisymmetry of $F_{ij}$.
\end{enumerate}
One may also consider $(\nabla_k F_{ij})(\nabla^k F^{ij})$, but this equals
$F_{ij}\nabla^2 F^{ij}$ plus a total spatial derivative and is therefore
cohomologous to a linear combination of the terms already listed. No other
independent gauge-invariant weight-$6$ structures exist at quadratic order.
Therefore, the cohomology at ghost number one is two-dimensional, spanned by the two terms in \eq{A-5D}, with independent coefficients
$c_T$ and $c_S$, confirming that these are the complete and genuine Lifshitz
scale anomalies for the gauge sector of a $(4+1)$-dimensional Lifshitz field
theory with $z = 2$.

\end{appendices}

\bibliographystyle{JHEP}

\bibliography{AC-ref} 
  
\end{document}